\documentclass[useAMS,usenatbib,twocolumn,usegraphicx]{mn2e}
\usepackage{times}

\newcommand{\beq}{\begin{equation}}
\newcommand{\eeq}{\end{equation}}
\newcommand{\beqar}{\begin{eqnarray}}
\newcommand{\eeqar}{\end{eqnarray}}

\newcommand{\mb}[1]{\mbox{\boldmath $#1$}}

\title[Alfv\'{e}n QPOs in magnetars]
{Alfv\'{e}n QPOs in magnetars in the anelastic approximation}

\author[Pablo Cerd\'a-Dur\'an, Nikolaos Stergioulas and Jos\'e A.~Font]
{Pablo Cerd\'a-Dur\'an$^1$\thanks{E-mail:cerda@mpa-garching.mpg.de},
Nikolaos Stergioulas$^2$\thanks{E-mail:niksterg@auth.gr},
and Jos\'e A.~Font$^3$\thanks{E-mail:j.antonio.font@uv.es}
\\
  $^1$Max-Planck-Institut f\"ur Astrophysik,
  Karl-Schwarzschild-Str.~1, 85741 Garching, Germany \\ 
  $^2$Department of Physics, Aristotle University of Thessaloniki,
  Thessaloniki 54124, Greece \\
  $^3$Departamento de Astronom\'{\i}a y Astrof\'{\i}sica,
  Universidad de Valencia, 46100 Burjassot (Valencia), Spain}

\begin{document}

\maketitle

\label{firstpage}

\begin{abstract}
	We perform two-dimensional simulations of Alfv\'{e}n oscillations in magnetars, 
	modeled as relativistic stars with a dipolar magnetic field. We use the 
	anelastic approximation to general relativistic magnetohydrodynamics,
	which allows for an effective suppression of fluid modes and an accurate description
	of Alfv\'{e}n waves. In addition, we compute Alfv\'{e}n oscillation frequencies
	along individual magnetic field lines with a semi-analytic approach, employing a 
        short-wavelength approximation. Our main findings are as follows: a) we 
	confirm the existence of two families of quasi-periodic oscillations (QPOs), 
	with harmonics at integer multiples of the fundamental frequency, as was found in 
	the linear study of Sotani, Kokkotas \& Stergioulas; b) the QPOs appearing 
	near the magnetic axis are split into two groups, depending on their symmetry across the
	equatorial plane. The antisymmetric QPOs have only odd integer-multiple harmonics;
	c) the continuum obtained with our semi-analytic approach
	agrees remarkably well with QPOs obtained via the two-dimensional simulations,
	allowing for a clear interpretation of the QPOs as corresponding to turning
	points of the continuum. This agreement will allow for a comprehensive study 
	of Alfv\'{e}n QPOs for a larger number of different models, without the need for
	time-consuming simulations. Finally, we construct empirical relations for the QPO 
	frequencies and compare them to observations of known Soft Gamma Repeaters.
	We find that, under the assumptions of our model and if the magnetic field of 
        magnetars is characterized by a strong dipolar component,
	and QPOs are produced near the magnetic pole, then one can place an upper limit to
        the mean surface strength of the magnetic field of about $3-8\times10^{15}$G.
\end{abstract}


\section{Introduction}
\label{sec:Intro}

The phenomenon of Soft Gamma Repeaters (SGRs) may allow us to determine fundamental properties of strongly magnetized, compact stars in the near future. SGRs produce giant flares with peak luminosities of $10^{44}$ -- $10^{46}$ erg/s, which display a decaying X-ray tail for several hundred seconds.  Up to now, three such giant flares have been detected, coming from SGR 0526-66 in 1979, SGR 1900+14 in 1998, and SGR 1806-20 in 2004.  Quasi-periodic oscillations (QPOs) have been observed in the X-ray tail of the latter two sources, following the initial discovery by \cite{Israel2005} (see \cite{WS2006} for a recent review). The timing analysis of these QPOs allowed the determination of their frequencies~\citep{WS2006}, which are approximately 18, 26, 30, 92, 150, 625, and 1840 Hz  in the SGR 1806-20 giant flare, and 28, 53, 84, and 155 Hz in the SGR 1900+14 giant flare. The frequency of many of these oscillations is similar to what one would expect for torsional modes of the solid crust of a compact star. This observation is in support of the proposal that SGRs are magnetars, i.e.~compact objects with very strong magnetic fields~\citep{DT1992}. During an SGR event, torsional oscillations in the solid crust of the star could be excited~\citep{Duncan1998}, leading to the observed frequencies in the X-ray tail. 

While a lot of theoretical work initially focused on the torsional oscillations of the solid crust (see 
\citet{Sotani2007a, SA2007} and references therein), 
more recently elasto-magnetic oscillations \citep{Levin2006,GSA2006,Levin2007,Lee2007, Lee2008} and pure Alfv\'en oscillations \citep{Sotani2008} have become attractive in the context of the observed oscillations in SGR giant flares. In particular,~\citet{Levin2006} and \cite{GSA2006} stressed the importance of crust-core coupling by a global magnetic field,  while  \cite{Levin2007} presented a toy-model of the Alfv\'en continuum and demonstrated that long-lived
QPOs can be produced at the edges or turning points of the continuum. In addition,~\cite{Sotani2008}, using two-dimensional linear simulations, found that in a magnetar model that uses 
ideal magneto-hydrodynamics (MHD) with a microphysical equation of state, long-lived QPOs appear near the magnetic axis and within the region of closed magnetic field lines. In addition, they found that an infinite number of ``overtone QPOs'' exist, with frequencies at integer multiples of the fundamental frequency, in remarkable agreement with observed QPO frequencies. Furthermore, \cite{Sotani2008} found that all Alfv\'en QPO frequencies are described by simple empirical relations, valid for any equation of state and magnetar mass, initiating the formulation of  magnetar asteroseismology based on global Alfv\'en QPOs (see also \cite{SA2007} for a formulation of asteroseismology based on torsional modes of the crust). 

The Alfv\'{e}n QPO model is attractive, since it reproduces the observed near-integer ratios of 30, 92 and 150 Hz in SGR 1806-20. Since these QPOs appear at odd-integer ratios, our current results suggest that they could correspond directly to the odd-parity QPOs appearing near the magnetic axis. In addition, the  Alfv\'{e}n QPO model is attractive for the reason that it is difficult to explain all of the frequencies of 18, 26 and 30 Hz in the SGR 1806-20 giant flare as being due to crustal shear modes, since the actual spacing of torsional oscillations of the crust is larger than the difference between the 26 and 30 Hz frequencies.  Only one of these frequencies could be the fundamental, $\ell=2,m=0$ torsional frequency of the crust, as the first overtone has a much higher frequency. Note that the 18 Hz frequency could be explained either by the Alfv\'en QPO model or as the fundamental crust
mode \citep{Steiner2009}. Similarly, the spacing between the 625Hz and a possible 720Hz QPO in SGR 1806-20 may be too small to be explained by consecutive overtones of crustal torsional oscillations. The
spectrum of observed QPOs may thus include both crustal normal-mode frequencies and Alfv\'{e}n QPO frequencies.

Here, we present the first two-dimensional numerical simulations of Alfv\'{e}n oscillations in magnetars in the {\it anelastic} approximation to general relativistic magnetohydrodynamics \citep{Bonazzola2007}, in which fluid modes are suppressed, obtaining efficient and accurate simulations of Alfv\'{e}n waves. In addition to the numerical simulations, we also compute Alfv\'{e}n oscillation frequencies along individual magnetic field lines with a semi-analytic approach, employing a 
short-wavelength approximation. The continuum obtained using standing-wave oscillations in the semi-analytic approach agrees remarkably well with QPOs obtained via our two-dimensional simulations, allowing for a clear interpretation of the QPOs as corresponding to turning points of the continuum. This agreement will allow for a comprehensive study of Alfv\'{e}n QPOs for a large number of different models, without the need of time-consuming simulations. In addition to the QPOs found in \cite{Sotani2008} (which were antisymmetric with 
respect to the equatorial plane) we also find a separate sub-family of QPOs that are symmetric. 
Furthermore, our implementation of the zero-traction boundary condition at the surface of the star shows that the QPOs generated near the axis have a nonzero amplitude at the surface, which would allow these oscillations to reach the magnetosphere. 

Our semi-analytic approach is not restricted to a dipolar field, but can be extended to other magnetic field configurations, while our  anelastic approximation does not suffer from the numerical instabilities encountered in the linear study of \cite{Sotani2008}. Furthermore, our anelastic 
approach could be used to extend the current sudy by including shear waves in the crust, which may prove to be important for the interpretation of observed QPOs. 

The paper is organized as follows: Section 2 describes the theoretical framework of the general relativistic MHD equations in the anelastic approximation. Section 3
discussed the equilibrium models of magnetars we consider. The Alfv\'en continuum and the conditions
for standing waves along individual field lines are studied with a semi-analytic approach in Section 4. In
Section 5 we present our axisymmetric simulations in the anelastic approximation, while in 
Section 6 we present empirical relations for the obtained QPO frequencies and compare them to 
observations of known SGR sources. Finally, we discuss our present results in Section 7. Unless stated otherwise, we adopt units of $c=G=1$, where $c$ and $G$ are the speed of light and the gravitational constant, respectively, and the spacetime metric signature is $(-,+,+,+)$. Latin (Greek) indices run from 1 to 3 (0 to 3) and boldface symbols indicate vectors.

 \section{The equations of general relativistic MHD in the anelastic approximation}
\label{seq:anelastic}
 
\begin{table*}
  \begin{minipage}{0.9\textwidth}
  \caption{Magnetar equilibrium models. From left to right the columns report the model name, the circumferential equatorial radius $R$, the coordinate equatorial radius $r_e$, the ratio of polar to equatorial coordinate radii (minus one), the gravitational mass $M$, the magnetic field strength at the pole $B_{\rm polar}$, the central rest mass density $\rho_{\rm c}$, the current density $j_0$, and the 
form of the current density in the interior (for model S1 the value of $j_0$ was not available). Model 
MNS2 (in bold) is our {\it reference model}.}
  \label{tab:ini_models}
  \begin{tabular}{lcccccccl}
    \hline
    Model & $R$ [km] & $r_e$ [km]& $r_p/r_e-1$ & $M$ [$M_\odot$] & $B_{\rm polar}$ [G] 
     & $\rho_{\rm c} [{\rm g\,cm^{-3}}]$& $j_0$ [A~m$^{-2}$] & current function\\
    \hline 
    MNS1  & $14.155$ & $11.998$ & $8\times10^{-6}$ & $1.40$ & $6.5\times 10^{14}$ 
    & $7.91\times 10^{14}$ &$2\times 10^{13}$ & \cite{Bocquet1995}\\
    {\bf MNS2}  & $14.157$ & $11.999$ & $8\times10^{-4}$   & $1.40$ & $6.5\times 10^{15}$ 
    & $7.91\times 10^{14}$ &$2\times 10^{14}$ & \cite{Bocquet1995}\\
    MNS3  & $14.168$ & $12.006$ & $5\times10^{-3}$    & $1.40$ & $1.6\times 10^{16}$ 
    & $7.91\times 10^{14}$ & $5\times 10^{14}$ & \cite{Bocquet1995}\\
    LMNS2 & $15.153$ & $13.322$ & $10^{-3}$   & $1.20$ & $5.1\times 10^{15}$ 
    & $5.47\times 10^{14}$ & $2\times 10^{14}$ & \cite{Bocquet1995}\\
    HMNS2 & $12.444$ & $9.941$ & $4\times10^{-4}$   & $1.60$ & $8.4\times 10^{15}$ 
    & $1.37\times 10^{15}$ & $2\times 10^{14}$ & \cite{Bocquet1995}\\
    \hline
    S1    & $14.153$ & $11.912$ & $0$      & $1.40$ & $4.0\times 10^{15}$ 
    & $7.91\times 10^{14}$ &        N/A          & \cite{Sotani2007a}\\
    \hline
  \end{tabular}
  \end{minipage}
\end{table*}

In our theoretical framework we use the 3+1 split of spacetime in which the metric reads
\begin{equation}
  ds^2 = - \alpha^2 \, dt^2 + \gamma_{ij} (dx^i + \beta^i \,dt)
  (dx^j + \beta^j \, dt),
\end{equation}
where $\alpha$ and $\beta^i$ are the lapse function and the shift vector, respectively. Furthermore, we adopt a conformally flat approximation in which the spatial 3-metric $\gamma_{ij}$ is
\begin{equation}
\gamma_{ij} = \phi^4 \hat{\gamma}_{ij},
\end{equation}
where $\phi$ is the conformal factor and $\hat{\gamma}_{ij}$ denotes the flat metric. The energy-momentum tensor of a magnetized perfect fluid is
\begin{equation}
  T^{\mu \nu} = (\rho h + b^2) \, u^\mu u^\nu +
  \left( P + \frac{b^2}{2} \right) g^{\mu \nu} - b^\mu b^\nu\,,
  \label{eq:tmunu_grmhd}
\end{equation}
where $\rho$ is the rest-mass density, $h$ is the specific enthalpy, $P$ is the isotropic fluid pressure, $u^{\mu}$ is the 4-velocity of the fluid, $b^{\mu}$ is the magnetic field measured by a comoving observer, and $g_{\mu\nu}$ is the 4-dimensional metric tensor. 

Following~\cite{anton06} we define the following set of conserved quantities 
(which are generalizations of the rest mass density, momentum density and energy density, 
respectively) 
\begin{eqnarray}
  D & = & \rho W,
  \label{eq:def:d}\\
  S_i & = & (\rho h + b^2) W^2 v_i - \alpha b_i b^0,
  \label{eq:def:si}\\
  \tau & = & (\rho h + b^2) W^2 - \left( P + \frac{b^2}{2} \right) -
  \alpha^2 (b^0)^2 - D,
  \label{eq:def:tau}	
\end{eqnarray}%
with $v^i$ being the 3-velocity of the fluid and $W\equiv\alpha u^0$ the Lorentz factor. This choice allows to cast the general-relativistic magneto-hydrodynamics (GRMHD) equations in the form of a flux-conservative hyperbolic system of equations, namely
\begin{equation}
  \frac{1}{\sqrt{- g}} \left[
  \frac{\partial \sqrt{\gamma} \mb{U}}{\partial t} +
  \frac{\partial \sqrt{- g} \mb{F}^i}{\partial x^i} \right] = \mb{S},
  \label{eq:hydro_conservation_equation}
\end{equation}
where $\sqrt{- g}=\alpha\sqrt{\gamma}$ and $\gamma=\det(\gamma_{ij})$ and where
the state vector, the flux vector, and the source vector are given, respectively, by
\begin{eqnarray}
  \mb{U} & = & [D, S_j, B^k],
  \label{eq:state_vector}
  \\
  \mb{F}^i & = & \Big[
  D \hat{v}^i, S_j \hat{v}^i + \delta^i_j \left( P + \frac{b^2}{2} \right) - \frac{b_j B^i}{W}, 
  \nonumber \\
  & &
  \hat{v}^i B^k - \hat{v}^k B^i \Big],
  \label{eq:flux_vector}
  \\
  \mb{S} & = & \left[ 0, \frac{1}{2} T^{\mu \nu}
  \frac{\partial g_{\mu \nu}}{\partial x^j}, 0, 0, 0  \right],
  \label{eq:source_vector}
\end{eqnarray}
where $\hat{v}^i = v^i - \beta^i/\alpha$.
In the above expressions, $B^i$ denotes the magnetic field components measured by an
Eulerian observer . The relation between $b^{\mu}$ and $B^{i}$ is \citep{anton06} 
\begin{equation}
b^{\mu} = \left [
\frac{W B^iv_i}{\alpha}, \frac{B^i + W^2 v^jB_j \hat{v}^i}{W}
\right ].
\end{equation}
Notice that we consider a barotropic equation of state (EOS), $P=K \rho^{\Gamma}$, for which an evolution equation for the energy $\tau$ is not required. This type of EOS is in particular appropriate for studying those small-amplitude stellar oscillations for which
thermal effects can be ignored. As shown in~\cite{anton06} the eigenvalues associated with the Jacobian matrices of the GRMHD system, $\lambda^i$, are an explicit function of the metric tensor, the fluid and
magnetic field variables and the speed of sound, $c_{\rm s}$, i.e.
\begin{equation}
\lambda^i = f (g_{\mu\nu}, \mb{U}, c_{\rm s}).
\end{equation}
The dependence on the speed of sound is a consequence of the pressure terms appearing in the fluxes which lead to the presence of a term $\partial P / \partial \mb{U} $ in the Jacobian matrices $\partial \mb{F}^i/ \partial \mb{U}$. For the case of a barotrope this term reads
\begin{equation}
\frac{\partial P}{\partial \mb{U}} = 
\frac{\partial P}{\partial \rho} 
\frac{\partial \rho}{\partial \mb{U}} 
= h c_{\rm s}^2
\frac{\partial \rho}{\partial \mb{U}}. 
\end{equation}
The main idea of the {\it anelastic approximation} \citep{Bonazzola2007} is to eliminate sound waves, in order to reduce the time-step restrictions of numerical codes that are based on time-explicit schemes. In the presence
of sound waves such restrictions are imposed by the dependence of the eigenvalues 
of the hyperbolic system on the speed of sound. Consider the following modified equations in which the pressure terms in the fluxes have been moved to the sources:
\begin{eqnarray}
  \mb{\tilde F}^i & = & \left[
  D \hat{v}^i, S_j \hat{v}^i + \delta^i_j \frac{b^2}{2} -
  \frac{b_j B^i}{W}, \hat{v}^i B^k - \hat{v}^k B^i \right],
  \label{eq:flux_vector_ana}
  \\
  \mb{\tilde S} & = & \left[ 0, \frac{1}{2} T^{\mu \nu}
  \frac{\partial g_{\mu \nu}}{\partial x^j} 
   -  \frac{1}{\sqrt{-g}} \frac{\partial \sqrt{-g} P}{\partial x^j}
, 0, 0, 0  \right].
  \label{eq:source_vector_ana}
\end{eqnarray}%
With this choice, the speed of sound does not appear in the eigenvalues of the corresponding Jacobian matrices. In fact, these modified eigenvalues can simply be obtained from the original eigenvalues 
by taking the limit of vanishing speed of sound, i.e.
\begin{equation}
\tilde \lambda^i (\gamma_{ij}, \mb{U}) 
= \lambda^i (\gamma_{ij}, \mb{U}, c_{\rm s}\to 0).
\end{equation}
With this modification the time-step required to satisfy the stability criterion imposed on 
hyperbolic equations by the Courant condition is much less restrictive, since it is not 
affected by sound waves. We note however, that this procedure cannot be applied in a more 
general context, since the source terms may become stiff due to the inclusion of the pressure, 
which may lead to unstable numerical evolutions of the system. Nevertheless, in the present 
anelastic approximation this approach is valid since we further assume that the pressure remains 
constant in time. In the case of a barotrope, this leads to \citep{Bonazzola2007}
\begin{equation}
\frac{\partial \sqrt{\gamma}D}{\partial t} = 0,
\end{equation}
which implies
\begin{equation}
\frac{1}{\sqrt{-g}} \frac{\partial \sqrt{-g} D \hat{v}^i}{\partial x^i} =0.
\label{eq:divsi}
\end{equation}
The anelastic approximation can be used in our numerical study since we deal with low-amplitude torsional  oscillations of an equilibrium star, which are of axial parity. Axial velocity perturbations couple only weakly to 
density perturbations, so that the density (and pressure) can be considered constant. 

In order to stabilize the  numerical code, we subtract from the source terms of the momentum equations
the initial equilibrium value of the flux-source balance (which should be zero but it is not due to discretization error). 
For axial oscillations at the linear level, only the perturbation in $B^\varphi$ and $S_{\varphi}$ are 
nonvanishing, so that we can keep $B^r$, $B^\theta$, $S_r$, and $S_{\theta}$ constant in time
(as long as the oscillations have small amplitude).

\section{Equilibrium models}
\label{sec:models}

In our analysis we use two different methods to build relativistic equilibrium models for non-rotating magnetars. The first method (valid up to moderate magnetic field strengths relevant for magnetars) relies on first computing a (nonmagnetized) Tolman-Oppenheimer-Volkoff (TOV) solution, and then separately computing the magnetic field structure. Since the TOV solution is described in Schwarzschild coordinates
we make the transformation to quasi-isotropic coordinates.
We further assume that the configuration of the magnetic field is that of a pure dipole. Details on the numerical method for constructing the magnetic field, as well as a representative figure of the magnetic field lines, can be found in~\cite{Sotani2007a}. The same configuration was used in~\cite{Sotani2008}. 

The second method we use for constructing equilibrium magnetar models is based on the  self-consistent solution of magnetized stars with a dipolar magnetic field, where all the effects of the magnetic field on the matter and spacetime are taken into account, presented by~\cite{Bocquet1995}.  These equilibrium models are computed using the {\tt LORENE} library~\footnote{http://www.lorene.obspm.fr}. We construct non-rotating polytropic equilibrium models with $\Gamma=2$ and $K=1.455\times 10^5$ (cgs units). The specific central enthalpy is chosen to be $h_{\rm c}=1.256$ and the form of the 
current density which gives rise to the magnetic field is given in  \cite{Bocquet1995}. Table~\ref{tab:ini_models} describes the main properties of all equilibrium initial models used in our study.
The equilibrium model MNS2 is used in the following Sections as a {\it reference model}, with a 
mass of 1.4$M_\odot$ and a magnetic field strength at the pole equal to  $B_{\rm pole}= 6.5\times 10^{15}$~G. 

\section{A Semi-Analytic Model}
\label{sec:alfven_oscillations}
 
The time evolution of the torsional Alfv\'en oscillations in the neutron star interior is a computationally expensive task, even in the anelastic approximation defined above.  This fact restricts the number of models that can be studied with our  numerical approach. Therefore, in addition to the numerical GRMHD simulations that will be presented in later Sections, we have developed a simplified approach for 
computing the oscillation spectrum. This semi-analytic model is based on the computation 
of Alfv\'en wave travel times along individual magnetic field lines (assuming a short-wavelength 
approximation) and requires only information on the initial equilibrium model and magnetic field configuration.

In the linear regime and in the limit of short wavelengths (compared to the length of magnetic field lines 
inside the star) torsional Alfv\'en oscillations can be treated as axial perturbations traveling strictly 
along individual magnetic field lines. This is evident in the local characteristic structure of the general relativistic MHD equations. The Alfv\'en eigenvalues of the system of equations for a given 
direction $\mb{ \hat n}$ is~\citep{anton06}
\begin{equation}
\lambda^{\boldmath{ \hat n}}_{\rm a\pm} = 
\frac{\mb{ b}\cdot\mb{ \hat n} \pm \sqrt{\mathcal C}\,\mb{ u}\cdot\mb{ \hat n} }
{b^0 \pm \sqrt{\mathcal C}\,u^0},
\end{equation}
where $\mathcal C \equiv \rho h + B^2$ and $\mb u$ is the fluid speed.  For a non-rotating initial 
equilibrium configuration $v^i=0$, and therefore the eigenvalue has a maximum when
 $\mb{ \hat n}$ is parallel to the magnetic field $\mb{ b}$. Then, we can construct the 
Alfv\'en velocity vector as
\begin{equation}
\mb{ v}_{\rm a} = \pm \frac{\alpha \mb{ b}}{\sqrt{\mathcal C}},
\end{equation}
which describes two possible waves along a magnetic field line, traveling in opposite directions. 

In this approximation, the path of an Alfv\'en wave inside the star can simply be found by 
integrating the equation
\begin{equation}
\frac{d \mb{ x}}{dt} = \mb{ v}_{\rm a} (\mb{ x}),
\label{eq:lineeq}
\end{equation}
starting from a given initial location $\mb{ x}_0$ inside the star. Since the result of 
this integration coincides with a magnetic field line passing through  $\mb{ x}_0$, 
we refer hereafter to the path of the Alfv\'en waves simply as magnetic field lines.  We integrate  (\ref{eq:lineeq}) numerically using a second order Runge-Kutta method.  Due to symmetry, 
only the part of the magnetic field lines in the upper plane is computed.The starting points 
for this integration are located in the equatorial plane, and the integration ends either at the surface 
of the star (for open magnetic field lines) or at another point in the equatorial plane (for close 
magnetic field lines). 

\begin{figure}
  \centering 
  \resizebox{0.45\textwidth}{!}{
   \hspace{-1.8 cm} \includegraphics*{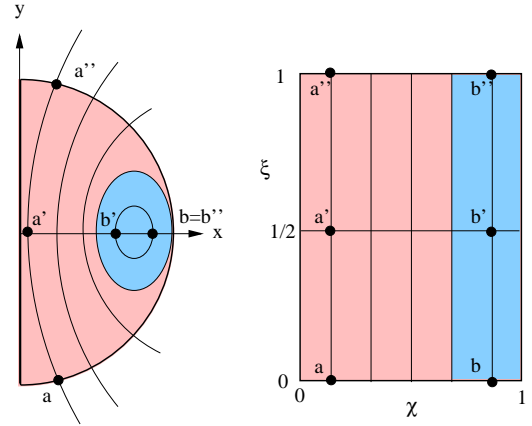}\hspace{0.2 cm}}
  \caption{Schematic illustration of the integration domain. 
In the region of open magnetic field lines (pink area) lines start at the
surface (point a), cross the equatorial plane (point a') and end at the opposite end of the 
surface (point a''). For closed magnetic field lines (blue area), points b and b'' coincide 
(left panel). The same magnetic field lines and points are shown in coordinates ($\chi, \xi$), 
which are adapted to Alfv\'en waves traveling along field lines (right panel, see text for
details).}
  \label{fig:sketch}
\end{figure}

Since the topology of the magnetic field lines considered here is rather simple, we can parametrize all magnetic field lines inside the star by their radial position $r$ in the equatorial plane, normalized to the 
radial position $r_c$ of the location in the equatorial plane where close magnetic field lines have zero
length (see Fig~\ref{fig:sketch}).  We call this parameter $\chi = r / r_c$, ranging from 0 to 1 
(notice that field lines ending in the region $r_c<r<r_e$ in the equatorial plane, where $r_e$ is 
the equatorial radius, originate in the region $r<r_c$). Using Alfv\'en waves that
travel along individual magnetic field lines as tracers (starting from the equatorial plane),   
the location of a specific point along a magnetic field line is a function of the parameter 
$\chi$ and of the arrival time $t_a$ of an Alfv\'en wave at this location i.e. $\mb{ x}=\mb{ x} (\chi;t_a)$. 
 Furthermore, we indicate as $t_{\rm tot}(\chi)$ twice the total travel time of an Alfv\'en wave traveling along a magnetic field line characterized by the parameter $\chi$, starting from the equatorial plane and ending either at the surface of the star or at another place in the equatorial plane.  
Then, the dimensionless parameter 
$\xi \equiv t_a(r,\theta;\chi)/t_{\rm tot}(\chi) - 1/2$, characterizes the location of points
 along an individual magnetic field line $\chi$.
For open magnetic field lines inside the star (see Fig. \ref{fig:sketch}), $\xi$ takes the value $0$ 
at the lower end of an open magnetic field line (i.e. at the surface of the star, below the equatorial
plane) and the value of 1 at the other end (at the surface of the star, above the equatorial
plane), while for closed field lines $\xi$ can take the value of 0 at any point and the value of
1 at the same point, after completing a full cycle. 
In this way, we can transform a 
magnetic field line from polar coordinates ($r,\theta$) to the dimensionless {\it magnetic-field-line-adapted coordinates} $(\chi, \xi)$. 
The above procedure could also be extended to more complicated or to three-dimensional topologies by adding more parameters and/or regions.

Next, assume that $Y(\xi,\chi;t)$ is a {\it displacement}
 that describes traveling waves  along a specific magnetic field
line $\chi$. Since we focus on individual field lines, we can simplify notation and
only show explicitly the dependence on $\xi$ and $t$ in what follows. For 
a traveling wave, $Y$ satisfies trivially the wave equation 
\begin{equation}
  \frac{\partial^2 Y (\xi, t)}{\partial t^2}
  = \sigma^2
  \frac{\partial^2 Y (\xi, t)}{\partial \xi^2},
  \label{eq:wave_equation}
\end{equation}
where $\sigma$ is the wave speed. It is easy to show that, owing to our use of the 
dimensionless arrival time $\xi=t_a/t_{\rm tot}$ for characterizing the location of points
along a magnetic field line, the wave speed $\sigma$ is constant and 
equal to $\sigma=1/t_{\rm tot}$. In other words, the dependence of the 
Alfv\'en speed on position has been absorbed into the definition of $\xi$.

To find a correspondence with QPOs, we are interested in solutions in the 
form of standing waves
\begin{equation}
  Y(\xi,t) = A(\xi) \cos(2 \pi f t + \varphi),
  \label{eq:standing_wave}
\end{equation}
where $f$ is the oscillation frequency and $\varphi$ is a phase. The spatial dependence can be written as
\begin{equation}
A(\xi) = a \, \sin(\kappa \xi + \phi),
\label{eq:spatial_perturb}
\end{equation}
where $\kappa$ is the wavenumber, $\phi$ a spatial phase, and $a$ the amplitude of the oscillation.  Substituting (\ref{eq:standing_wave}) into (\ref{eq:wave_equation}), the resulting dispersion relation is
simply
\begin{equation}
  f = \frac{\sigma \kappa}{2\pi}.
  \label{eq:dispersion_rel}
\end{equation}
If we assume the existence of standing waves for a single magnetic field line, then $Y(\xi,t)=Y(\xi,t+t_{\rm cycle})$, where $t_{\rm  cycle}$ is the time it takes for an Alfv\'en wave to complete a full cycle along 
the magnetic field line and return to its initial position (for open magnetic field lines, the wave will
be reflected at the surface above and below the equatorial plane before completing the cycle). 
An infinite number of discrete oscillation frequencies corresponds to standing waves
\begin{equation}
  f_n = \frac{n}{t_{\rm cycle}}, \qquad n= 1, 2, 3 ...
\end{equation}
with $f_1=1/t_{\rm cycle}$ being the frequency of the fundamental mode. 

To simplify further the analysis, we decompose the spatial 
part of the perturbations into odd (-) and even (+) parity 
components with respect to the midpoint $\xi=1/2$
\begin{equation}
  A_n (\xi) \equiv A_n^{+}(\xi) +  A_n^{-}(\xi).
\end{equation}
If the background equilibrium (unperturbed) model is symmetric with respect to the equatorial plane, then $\xi=1/2$ coincides with the equatorial plane, and the two components of the perturbation are just symmetric (+) and antisymmetric (-) oscillations with respect to the equatorial plane. Note however, that 
such a split would also exist in a general case, without equatorial plane symmetry, since there is always a midpoint $\xi=1/2$.  The symmetry of the oscillations fixes the spatial phase of each component, so that
\begin{eqnarray}
  A_n^{+}(\xi) &=& a^+_n \, \cos [\kappa_n (\xi - 1/2)], \nonumber \\
  A_n^{-}(\xi) &=& a^-_n \, \sin [\kappa_n (\xi - 1/2)].
  \label{eq:sadecomposition}
\end{eqnarray}
This decomposition can be related to Eq.~(\ref{eq:spatial_perturb}) for each frequency $f_n$, considering
\begin{eqnarray}
  {a_n}^2  &=& {a_n^{+}}^{\,2} +  {a_n^{-}}^{\,2}, \nonumber \\
  \phi_n &=& \arctan \left(  a^-_n / a^+_n \right ) - \kappa_n / 2. 
\end{eqnarray}

To completely determine the solution of the problem one has to further consider the topology of the magnetic field lines and the boundary conditions. Due to the solenoidal condition satisfied by the magnetic field, only two possible magnetic field line topologies are possible for any field inside the star: open or closed field lines, which we consider separately, next.

\subsection{Closed magnetic field lines inside the star}
 
We consider first magnetic field lines that close inside the star. In this case, standing-wave 
perturbations fulfill spatial periodicity
\begin{equation}
  Y(1, t) = Y (0, t).
\end{equation}
Therefore, the time that takes for a perturbation to complete a cycle is just the time to travel along the line, $t_{\rm cycle}=t_{\rm  tot}$. The frequency of each standing wave is
\begin{equation}
f_n = \frac{n}{t_{\rm tot}} \quad ;\quad  n = 1, 2, 3 ...
\end{equation}

Using the dispersion relation (\ref{eq:dispersion_rel}) the wave number for each standing wave can be
found as
\begin{equation}
\kappa_n = 2 \pi \,n \qquad  n = 1, 2, 3, ...
\end{equation}
Therefore, only standing waves with discrete frequencies $f_n$ are possible along a single magnetic field line. The number of nodes for each standing wave is just $2\,n$, and, using (\ref{eq:sadecomposition}) the symmetric and antisymmetric component of the oscillations are
\begin{eqnarray}
  A_n^{+}(\xi) &=& a^+_n \, \cos (2 \pi n \,\xi), \nonumber \\
  A_n^{-}(\xi) &=& a^-_n \, \sin (2 \pi n \,\xi).
\end{eqnarray}
If we consider purely symmetric oscillations, then $a_n^- = 0$ and therefore the spatial phase is fixed to $\phi_n=0$, for every $n$. In the case of purely antisymmetric oscillations $a_n^+ = 0$ and hence, $\phi_n=\pi/2$. {\it In the general case, without any symmetry in the perturbations, the spatial phase depends on the initial condition and can be different for each $n$.}

Since in this region of closed field lines each magnetic field line is decoupled from the others, the frequency $f(\xi)$ forms a {\it continuum} for each overtone. 

 \subsection{Open magnetic field lines}
 
In this case, each magnetic field line crosses the surface of the star at two points, $\xi=0$ and $\xi=1$. The time it takes an Alfv\'en wave to complete a cycle, i.e. to travel along the magnetic field line and return back to the starting point is $t_{\rm cycle}=2\,t_{\rm tot}$. Therefore, the allowed frequencies for standing waves are
\begin{equation}
  f_n = \frac{n}{2 t_{\rm tot}} \quad;\quad n = 1, 2, 3 ...
\end{equation}
and therefore
\begin{equation}
  \kappa_n = \pi \, n \quad;\quad n = 1, 2, 3 ...
\end{equation}
The symmetric and antisymmetric components read in this case
\begin{eqnarray}
  A_n^{+} (\xi) &=& a_n^+\, \sin \left (\pi\, n \xi \right ) \quad;\quad n = 1, 3, 5 ...
  \label{eq:upqo_sym_odd}\\
  A_n^{+} (\xi) &=& a_n^+\, \cos \left (\pi\, n \xi \right ) \quad;\quad n = 2, 4, 6 ...
  \label{eq:upqo_sym_even}\\
  A_n^{-} (\xi) &=& a_n^-\, \cos \left (\pi\, n \xi \right ) \quad;\quad n = 1, 3, 5 ...
  \label{eq:upqo_antisym_odd}\\
  A_n^{-} (\xi) &=& a_n^-\, \sin \left (\pi\, n \xi \right ) \quad;\quad n = 2, 4, 6 ...
  \label{eq:upqo_antisym_even}
\end{eqnarray}
It is important to note that symmetric oscillations with even $n$ and antisymmetric oscillations with odd $n$, i.e.~Eqs.~(\ref{eq:upqo_sym_even}) and~(\ref{eq:upqo_antisym_odd}), fulfill the conditions  $A_n (0)=A_n (1)=0$ at the boundaries. On the other hand symmetric oscillations with odd $n$ and antisymmetric oscillations with even $n$ i.e.~Eqs.~(\ref{eq:upqo_sym_odd}) and~(\ref{eq:upqo_antisym_even}) satisfy the conditions  $\partial_\xi A_n (0)= \partial_\xi A_n (1)=0$ at the boundaries. Therefore, the possible standing waves depend on the symmetry of the perturbation with respect to the equatorial plane and on the boundary conditions imposed at the surface.

\subsection{Boundary conditions at the surface}
\label{sec:bc}

The boundary conditions at the surface of the star should ensure the continuity of traction.
In the present approach, we neglect the dynamics in the magnetosphere, so that the traction 
is zero at the surface. If we do not consider the presence of a crust, the zero traction
condition for Alfv\'en oscillations at the surface would lead to the boundary condition that 
the gradient of $Y$ along each field line should vanish at the surface. However, in reality 
there exist a solid crust in the star, which occupies a region of about 5 to 10\% of the radius.
This changes the boundary condition at the surface of the star. Furthermore, the oscillations
behave differently in the fluid region, where only Alfv\'en waves are possible, and in the crust,
where both Alfv\'en and shear waves appear. In the case that the magnetic field is not too 
strong, the Alfv\'en speed in the crust is comparable or smaller to the shear speed. In this
case we can make the approximation that the boundary condition at the surface of the star 
corresponds to the zero-traction condition for shear waves in a nonmagnetized star, i.e.
\begin{equation}
  \frac{\partial Y}{\partial r} = 0.
\label{boundarycondition}
\end{equation}
Furthermore, we neglect the shear waves in the crust. This approximation is valid in two cases. 
First, if shear and Alfv\'en waves have similar velocities, then the velocity of the perturbations
in the crust is underestimated (but correct in order of magnitude), and the direction of 
propagation is not exactly along the magnetic field line. 
Second, if shear waves dominate the dispersion relation at the crust (low magnetic field),
then the perturbations traveling along magnetic field lines inside the core and reaching the 
crust interface, will propagate almost immediately inside the crust (in comparison with the 
travel time inside the core) to reach the surface where the boundary condition assumed in
this work holds. Therefore our approximation in this limit has the meaning of considering
a very thin crust, and hence neglecting its inertia. In both of the cases, the approximation can produce small quantitative 
differences in the frequency spectrum and damping time, but the qualitative behavior should
not change. This approximation should be valid as long as the magnetic field is not too strong 
(possibly larger than $10^{16}$G). This boundary condition, contrary to the case 
in which the presence of a crust is completely neglected, introduces a coupling between the
different magnetic field lines which affects the time evolution of the QPOs that appear 
at turning points.

Since the perturbation is a function $Y(\xi, \chi; t)$ (in the following we 
include again the explicit dependence on $\chi$) the boundary condition
 (\ref{boundarycondition}) becomes
\begin{equation}
  \frac{\partial Y } {\partial \xi} \frac{\partial \xi}{\partial r}
  + \frac{\partial Y } {\partial \chi} \frac{\partial \chi}{\partial r} = 0.
\end{equation}
Using the standing-wave solution (\ref{eq:standing_wave}) yields 
\begin{eqnarray}
\frac{\partial A (\xi, \chi)}{\partial \xi} \cos (2\pi f\, t + \varphi)
\frac{\partial \xi}{\partial r} 
+ \left [
\frac{\partial A (\xi, \chi)}{\partial \chi} \cos (2\pi f\, t + \varphi) 
\right . & & \nonumber \\ \left .
  - A (\xi,\chi) \sin  (2\pi f\, t + \varphi) 
  \left ( 2 \pi t \frac{\partial f}{\partial \chi} 
+ \frac{\partial \varphi}{\partial \chi} \right ) 
\right ]
\frac{\partial \chi}{\partial r}
= 0. &&
\label{eq:bc_equation}
\end{eqnarray}
Due to these boundary conditions at the surface, standing-wave solutions are
possible in two different cases. For perturbations which vanish at the surface, 
i.e. for solutions of the form  (\ref{eq:upqo_sym_even}) and~(\ref{eq:upqo_antisym_odd}), 
the  only solution of (\ref{eq:bc_equation}) is the trivial solution for vanishing amplitude. 
On the other hand, for solutions of the form  (\ref{eq:upqo_sym_odd}) 
and~(\ref{eq:upqo_antisym_even}), (satisfying $\partial_\xi A_n=0$ at the surface) 
a standing-wave solution exists if 
\begin{equation}
  \left ( 2 \pi t \frac{\partial f}{\partial \chi} 
+ \frac{\partial \varphi}{\partial \chi} \right ) 
\frac{\partial \chi}{\partial r} = 0,
\quad\quad
\frac{\partial a}{\partial \chi} 
\frac{\partial \chi}{\partial r} = 0.
\end{equation}
These conditions can be fulfilled only in two cases. First, if
\begin{equation}
 \frac{\partial f}{\partial \chi} =0,
 \quad  \quad
 \frac{\partial \varphi}{\partial \chi} =0,
 \quad  \quad
 \frac{\partial a}{\partial \chi} =0,
\end{equation}
i.e.  if $f(\chi)$, $\varphi(\chi)$ and $a(\chi)$ have a common turning point, or if 
they are constant. The second case is when
\begin{equation}
\frac{\partial \chi}{\partial r} = 0,
\end{equation}
i.e. for those magnetic field lines that cross the stellar surface perpendicular to it. 
In axisymmetry, this always holds for the magnetic field line along the pole. Therefore, 
only at the magnetic pole an exact standing wave solution is possible. The spatial 
part of the solution is
\begin{equation}
  A_n (\xi) = a_n \, \cos (\pi n \,\xi),
\end{equation}
which has $n$ nodes inside the star. As we discuss next, one can still define a standing
wave along any magnetic field line, on a short enough timescale.

\begin{figure}
  \centering
  \resizebox{0.45\textwidth}{!}{\includegraphics*{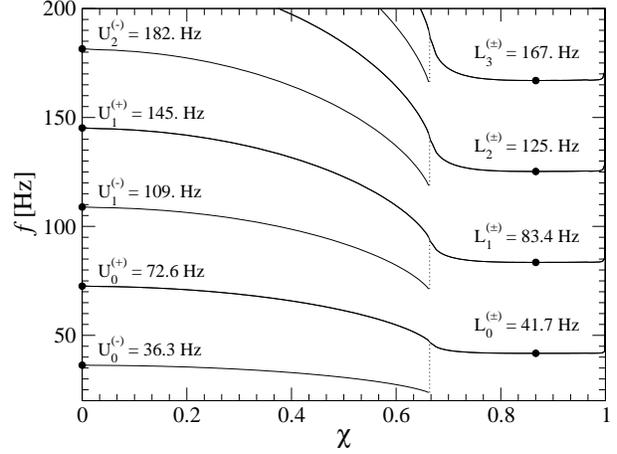}}
  \caption{Standing-wave continuum and QPOs for the reference model MNS2. 
Filled circles  denote the (upper and lower) QPOs. See text for details.}
  \label{fig:mns2cont}
\end{figure}

\subsection{Standing-wave continuum and QPOs}

According to the standing-wave solution for the region of open field lines, presented in
the previous section, as long as 
\begin{equation}
2 \pi \frac{\partial f}{\partial \chi} t \ll1,
 \quad  \quad
 \frac{\partial \varphi}{\partial \chi} \sim 0,
 \quad  \quad
 \frac{\partial a}{\partial \chi} \sim 0.
\label{conditions}
\end{equation}
the time-evolution of an initial perturbation will still be close to a standing wave. 
Therefore, the characteristic timescale on which phase mixing occurs and a 
standing wave is damped, is 
\begin{equation}
\tau_{\rm d} = \frac{1}{2 \pi} \left ( \frac{\partial f}{\partial \chi} \right )^{-1}.
\label{eq:damping}
\end{equation}
Consequently, on a timescale smaller than $\tau_{\rm d}$, 
one can define a {\it standing-wave continuum} for magnetic field lines throughout 
the star, for each overtone.

In Fig.~\ref{fig:mns2cont} we display the different branches of the standing-wave 
continuum of frequencies for the reference model MNS2, in a frequency vs. magnetic
field line plot. The continuum can be separated into two parts, 
one for the region of open field lines ($0<\chi < 0.66$) and another for the region 
of closed field lines ($0.66<\chi< 1.0$). 
Standing waves in the region of open field lines can be divided into two sub-families,
according to their symmetry with respect to the equatorial plane, with distinct frequencies. 
In the region of closed field lines, these two sub-families have identical frequencies. 
 Notice that, for each overtone, the branch which corresponds to 
symmetric standing waves in the region of open field lines smoothly joins to the two
overlapping branches of symmetric and antisymmetric standing waves in the region
of closed magnetic field lines.

For the particular magnetic field configuration used here, the standing-wave 
frequencies have a local maximum at $\chi=0$ (i.e. at the pole). 
Notice that conditions (\ref{conditions}) are always fulfilled at the pole, since it is
a turning point of the continuum (due to axisymmetry). Hence, near the pole, we
expect local standing waves to give rise to long-lived QPOs. The predicted 
frequencies of these QPOs at the  pole (the ``upper'' QPOs) 
are shown in Fig.~(\ref{fig:mns2cont}) and
labeled as $U^{(+)}_n$ and $U^{(-)}_n$, for symmetric and antisymmetric 
standing waves, respectively. Here, the index $n$ labels the different overtones 
(starting with the fundamental frequency, $n=0$) and also coincides with the number of
nodes appearing between center and pole.

In the region of closed field lines, the standing-wave continuum for each overtone 
features a local minimum. Notice that the continuum for the most part of this region is nearly 
flat\footnote{At $\chi=1$ the frequency tends to infinity, because the length of the magnetic
field line tends to zero.}, 
implying a slow rate of phase-mixing. These local minima also give rise to long-lived 
QPOs in our numerical simulations (the ``lower'' QPOs), which are labeled 
as $L^{(\pm)}_n$ (the frequency of symmetric and antisymmetric standing waves coincide). Here,  
the index $n$ again labels the different overtones (starting with the fundamental frequency, $n=0$).
For the fundamental frequency, two nodes appear along a closed field line, corresponding
to a single nodal line crossing the region of closed field lines. For each consecutive overtone, 
our semi-analytic model predicts an additional nodal line. 

\begin{table}
  \begin{minipage}{0.43\textwidth}
  \caption{QPO frequencies for the reference model MNS2, in the anelastic 
approximation. In parenthesis we give
    the relative difference with respect to the semi-analytic approach. The frequency resolution of
the numerical simulations is $1.25$~Hz.}
  \label{tab:freqs}
  \begin{tabular}{ccccccccccc}
    \hline
    ``upper'' QPOs       &  f (Hz)            & ``lower QPOs''  &  f (Hz)      \\
    \hline 
    $U^{(-)}_0$ & 35.1 (4.\%)  & $L^{\pm}_0$ & 41.30 (1.0\%) \\[0.3 em]
    $U^{(+)}_0$ & 72.6 (0.05\%)& $L^{\pm}_1$ & 83.85 (0.6\%) \\[0.3 em]
    $U^{(-)}_1$ & 110.2(1.1\%) & $L^{\pm}_2$ & 126.4 (1.0\%) \\[0.3 em]
    $U^{(+)}_1$ & 147.8(1.8\%) & $L^{\pm}_3$ & 170.2 (2.0\%) \\[0.3 em]
    $U^{(-)}_2$ & 182.8(0.7\%) & $L^{\pm}_4$ & 210.3 (0.9\%) \\[0.3 em]
    $U^{(+)}_2$ & 220.4(1.2\%) & $L^{\pm}_5$ & 251.6 (0.6\%) \\[0.3 em]
    $U^{(-)}_3$ & 256.7(1.0\%) & $L^{\pm}_6$ & 295.4 (1.2\%) \\[0.3 em]
    $U^{(+)}_3$ & 294.3(1.3\%) & $L^{\pm}_7$ & 336.7 (0.9\%) \\[0.3 em]  
    \hline
  \end{tabular}
  \end{minipage}
\end{table} 

\begin{figure*}
  \includegraphics[height=0.69\textheight]{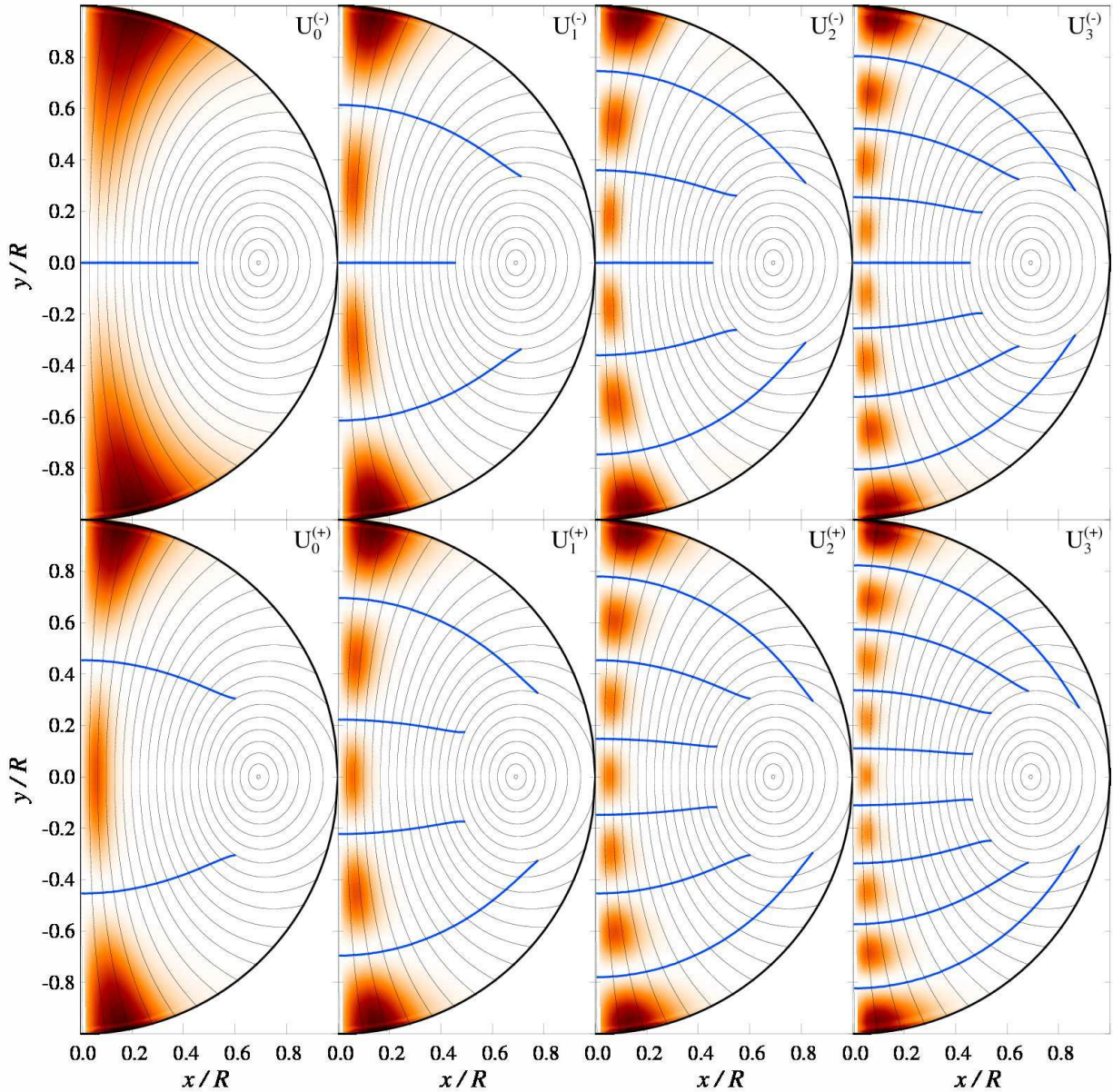}
  \caption{Spatial pattern of the effective amplitude for several upper QPOs, 
obtained in the anelastic approximation for the reference model MNS2. 
Both antisymmetric (upper panels) and 
symmetric (lower panels) oscillations are shown. 
The thin black lines are  magnetic field lines, while the thick black line is the stellar surface.
The nearly horizontal (blue) lines are the nodal lines predicted by the semi-analytic model for
standing-wave oscillations in the region of open field lines.}
  \label{fig:mns2uqpo}
\end{figure*}

\begin{figure*}
  \includegraphics[height=0.39\textheight]{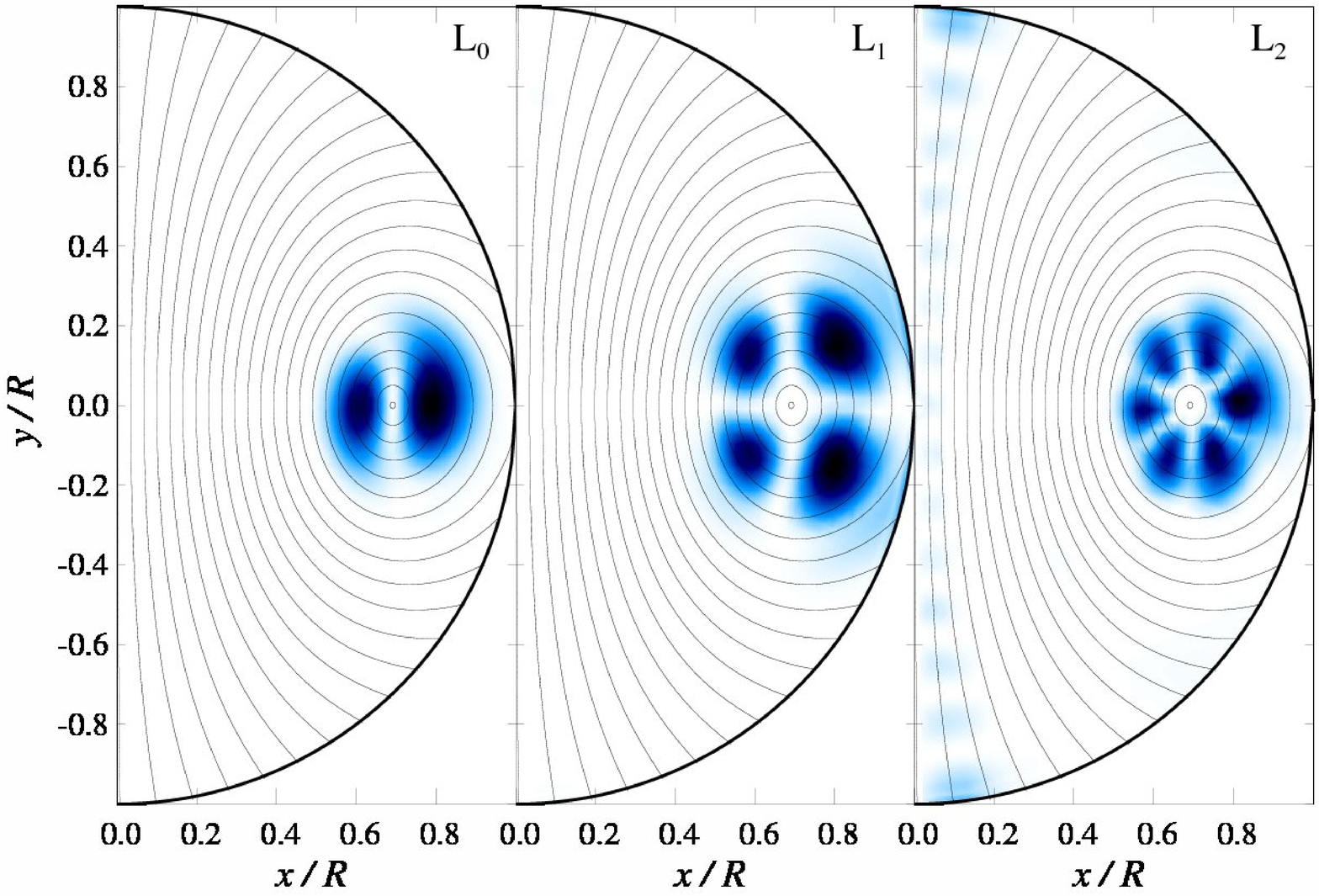}
  \caption{Spatial pattern of the lower QPOs (reference model). Color coded is the oscillation amplitude, $\dot {\mathcal Y}(r,\theta; f)$, for the fundamental lower QPO ($L_0$) and the first two overtones ($L_1$ and $L_2$). The thin black lines show the magnetic field line structure inside the star and the thick black line represents 
the stellar surface.}
  \label{fig:mns2lqpo}
\end{figure*}

\section{Numerical simulations}

Our second approach in studying Alfv\'en QPOs in magnetars involves two-dimensional 
numerical simulations using the nonlinear GRMHD code described in \citet{CerdaDuran2008}, in
which we implemented the anelastic approximation presented in Sec. \ref{seq:anelastic}.
We use spherical polar coordinates under the assumption of axisymmetry. This code 
solves the GRMHD equations coupled to the Einstein equations for the evolution of a dynamical 
spacetime (in the approximation of spatial conformal flatness) and has been developed with the main objective of studying astrophysical scenarios in which both high magnetic fields and strong 
gravitational fields are involved, such as the magneto-rotational collapse of stellar cores, the collapsar 
model of gamma-ray bursts, and the evolution of magnetized neutron stars.  The code is based on high-resolution shock-capturing schemes for solving the GRMHD equations, which are cast in a first-order, flux-conservative hyperbolic form, supplemented by the flux constraint transport method to ensure the solenoidal condition of the magnetic field.  In addition, the code can handle several EOS, from simple analytical expressions to microphysical tabulated ones. The robustness of the code was demonstrated in ~\citet{CerdaDuran2008} using a number of stringent tests, such as relativistic shocks, highly magnetized fluids, equilibrium configurations of magnetized neutron stars, and the magneto-rotational core collapse of a realistic progenitor. Although we use a nonlinear 
numerical code, we restrict attention to small perturbations and focus only on the linear 
behaviour of the oscillations. 

The fundamental variable in our simulations is the velocity component $v^\phi$ and 
the corresponding {\it displacement} $Y$, which we define through the relation
\begin{equation}
\dot Y = \alpha v^\phi,
\end{equation}
where a dot denotes a time derivative. 
Following the presentation of our anelastic approximation in Sec.  \ref{seq:anelastic}, we
summarize here the particular assumptions made in our numerical implementation:
\begin{enumerate}
\item We assume a fixed spacetime metric (Cowling approximation). Since we consider 
nonrotating, stationary configurations, the shift vector vanishes, $\beta^i=0$, 
for the gauge choice made in the spatial conformal flatness approximation. 
\item The evolution of the continuity equation is not performed, since this is the main assumption of the anelastic approximation.
\item Since we restrict our study to small-amplitude oscillations, the $v_r$and  $v_\theta$ components
of the 3-velocity couple only weakly to the dominant component $v_\varphi$ and can therefore be
neglected. 
\item In axisymmetry and with $v_r=v_{\theta}=0$, the poloidal part of the magnetic field, $B_r$ and $B_{\theta}$, does not evolve in time and can be kept constant throughout the simulation. Thus, 
only $B_\phi$ is evolved in time and describes torsional oscillations (axial parity).
\item The combination of the previous two assumptions makes it possible to avoid the evolution of the momentum equations for $S_r$ and $S_{\theta}$. This approach simplifies the requirement given by Eq.~(\ref{eq:divsi}), as it is trivially fulfilled in our case. 
\item The boundary conditions at the surface of the star are assumed to be the zero-traction
conditions for shear waves in a solid crust (in order to mimic the coupling of different magnetic
field lines by a solid crust, see Sec. \ref{sec:bc}).  For $v^\varphi$ this corresponds to 
$\partial_r \dot Y = \partial_r (\alpha  v^\varphi) = 0$, which is equivalent 
to $\partial_r Y = 0$ if the initial conditions also fulfill  that $\partial_r Y (t=0)= 0$.
 The boundary condition for $B^{\varphi}$ is computed numerically from the condition 
for $v^\varphi$ by means of  the induction equation. 
\end{enumerate}

With all these assumptions only the equations for $S_\varphi$ and $B_\varphi$ are evolved in our anelastic simulations of torsional Alfv\'en modes. We choose an initial perturbation of the form
\begin{equation}
  Y (t=0) = 0 \quad ; \quad  \dot Y (t=0)= f(r) \, b(\theta),
  \label{eq:ini_pert}
\end{equation}
where 
\begin{eqnarray}
  f (r) &=& \sin \left ( \frac{3 \pi}{2} \frac{r}{r_e}\right), \nonumber \\
  b ( \theta ) &=& b_2 \frac{1}{3} \partial_\theta  P_2 (\cos{\theta})
  +  b_3 \frac{1}{6} \partial_\theta  P_3 (\cos{\theta}),
\end{eqnarray}
where $P_2$ and $P_3$ are the corresponding Legendre polynomials. The coefficients $b_2$ and $b_3$ are such that the maximum initial amplitude of $\dot Y$ throughout the star is just $b_2+b_3$. By setting $b_2=0$ the perturbation is symmetric with respect to the equator, and by setting $b_3=0$ it is antisymmetric. Any other combination has no defined symmetry. 
\begin{figure}
  \includegraphics[width=0.45\textwidth]{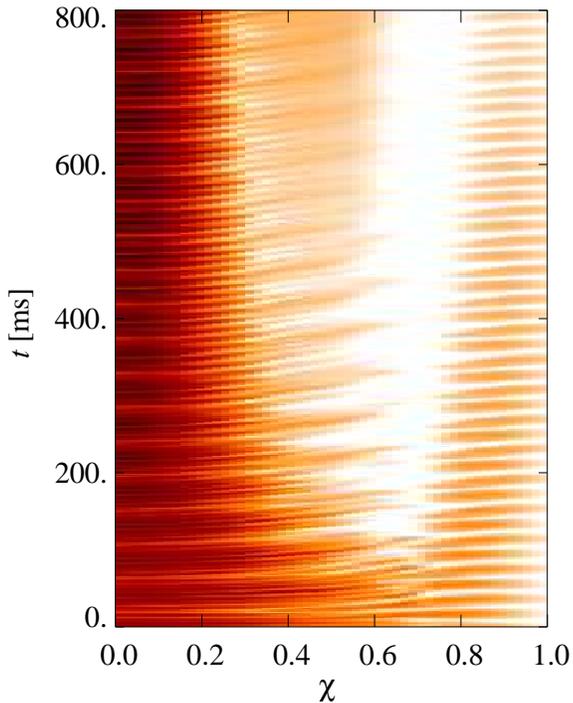}
  \caption{ Time evolution of the maximum perturbation amplitude, $\dot Y_{\rm max} (\chi; t)$ (color plotted, logarithmic scale), for each magnetic field line, $\chi$, of the reference model with a resolution of $80\times80$. The color scale ranges from zero (white) to its  maximum value (red-black).}
  \label{fig:mns2evolution}
\end{figure}

It is worth noting here that the zero traction boundary condition implies a non-zero angular momentum flux at the surface, since $F^r(S_\varphi) = - b_\varphi B^r / W $.  Because of this, the star acquires
a net angular momentum. This effect is just an artifact, arising from neglecting the presence
of a magnetosphere, where part of the Alfv\'en wave energy would be transmitted (the actual
boundary conditions would then be the continuity of a nonvanishing traction across the
surface). Notice that, antisymmetric perturbations add no net angular momentum to the system, while symmetric perturbations introduce some amount. For small enough oscillation amplitudes (such as those considered here) the net angular momentum added has a negligible effect on  the evolution of torsional modes. 

Indeed, the radial function $f(r)$ is chosen to satisfy the boundary conditions at the surface of the star and to minimize the net angular momentum added by the perturbation. We have checked that the net angular momentum added by this particular choice is much smaller than the angular momentum introduced by the boundary conditions during the evolution. Furthermore, we note that due to particular choice 
(\ref{eq:ini_pert}) for the initial perturbation and since the equilibrium magnetic field is poloidal, the
initial value for $B^\varphi$ is $B^\varphi (t=0) = 0$.

\subsection{Simulations for the reference model}

We begin our simulations by studying the evolution of the reference model MNS2. To this equilibrium model we add an initial velocity perturbation, consistent with Eq.~(\ref{eq:ini_pert}), with $a_2=a_3=5\times 10^{-6}$. This perturbation has no defined symmetry and its amplitude is small enough to be
considered at the linear level. By adding this perturbation torsional oscillations are induced in the star which can be observed as oscillations of $v^\varphi$ and $B^\varphi$ in time.

\begin{figure*}
  \includegraphics[height=0.5\textwidth]{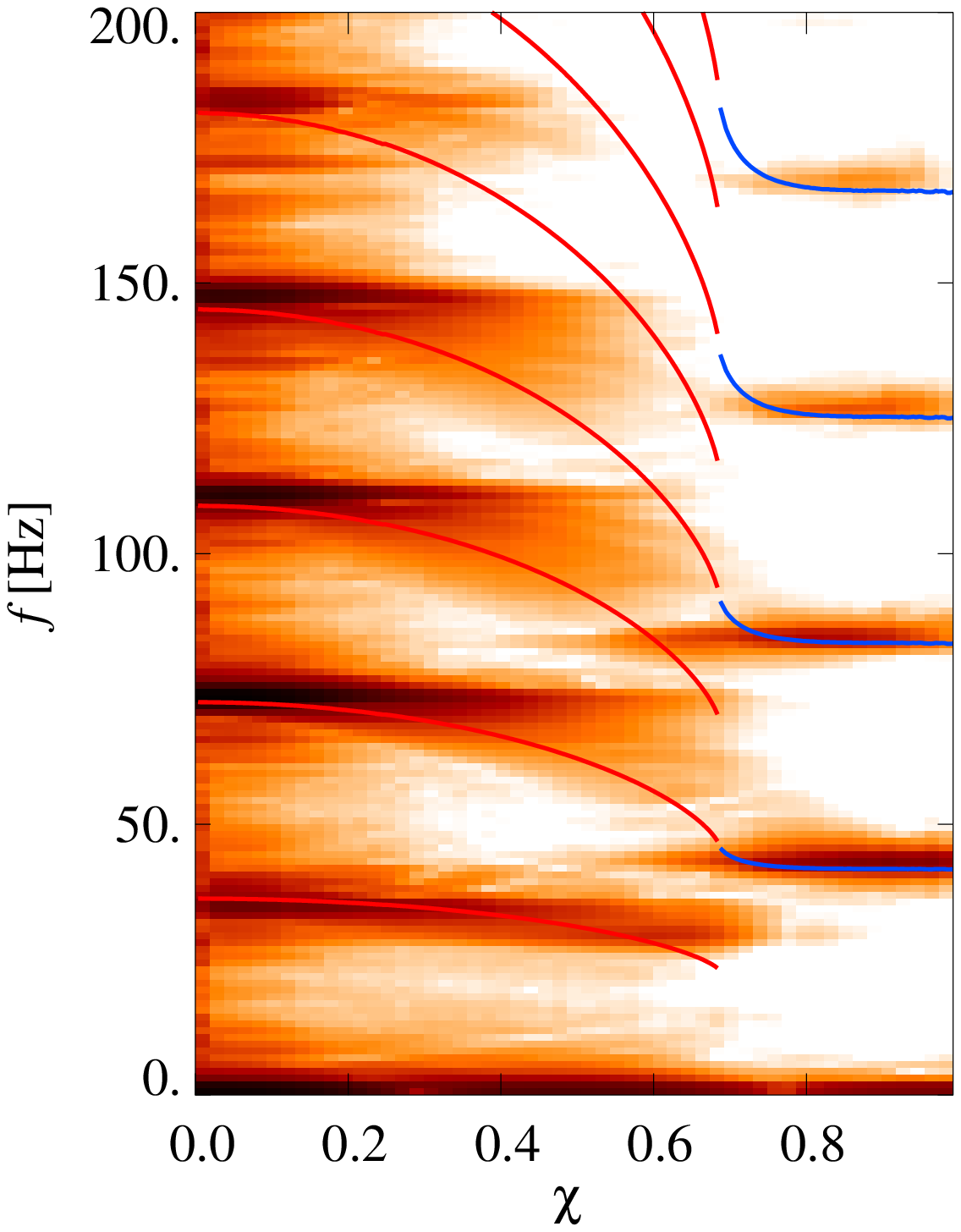}
  \hspace{-0.14cm}
  \includegraphics[height=0.5\textwidth]{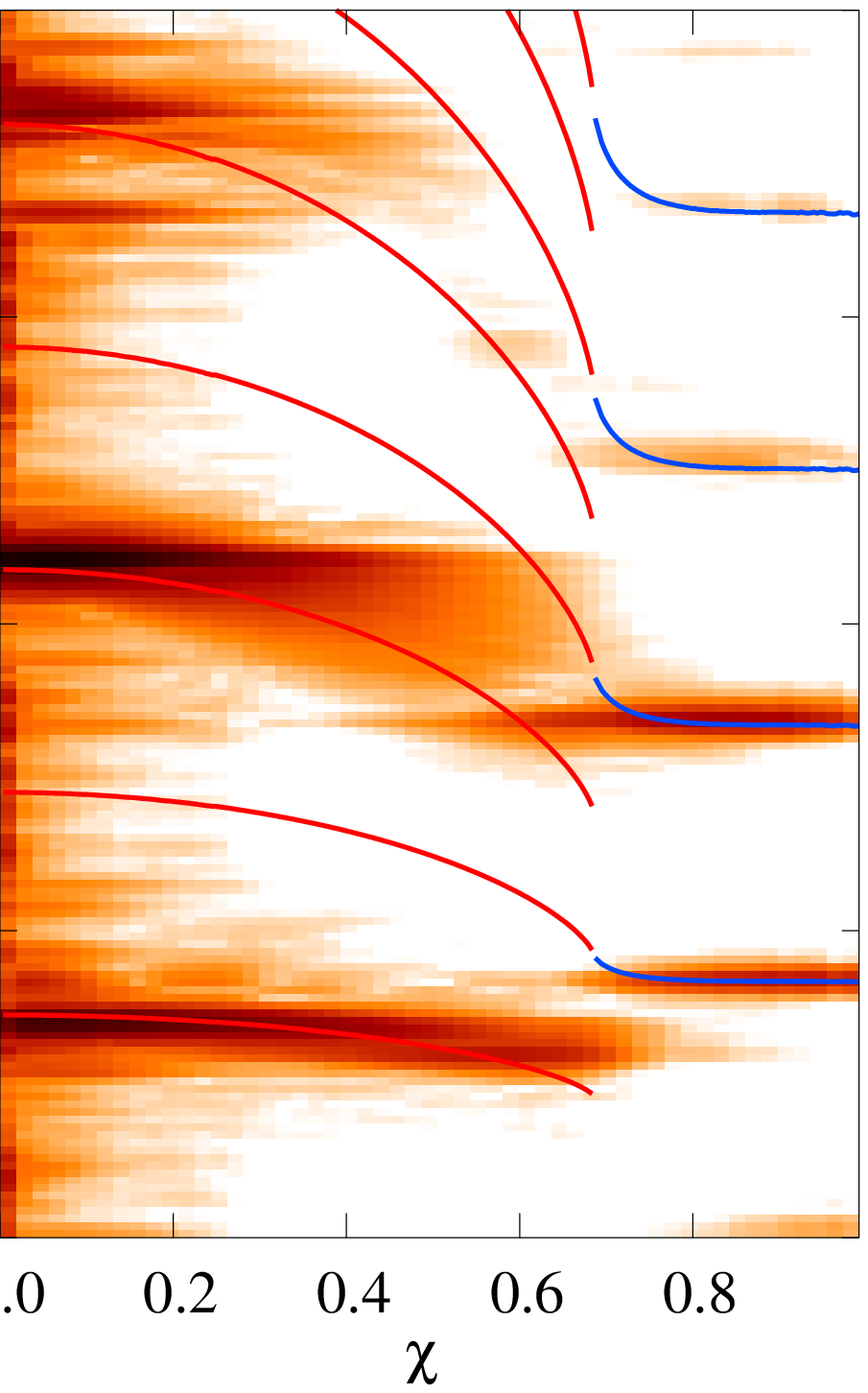}
  \hspace{-0.14cm}
  \includegraphics[height=0.5\textwidth]{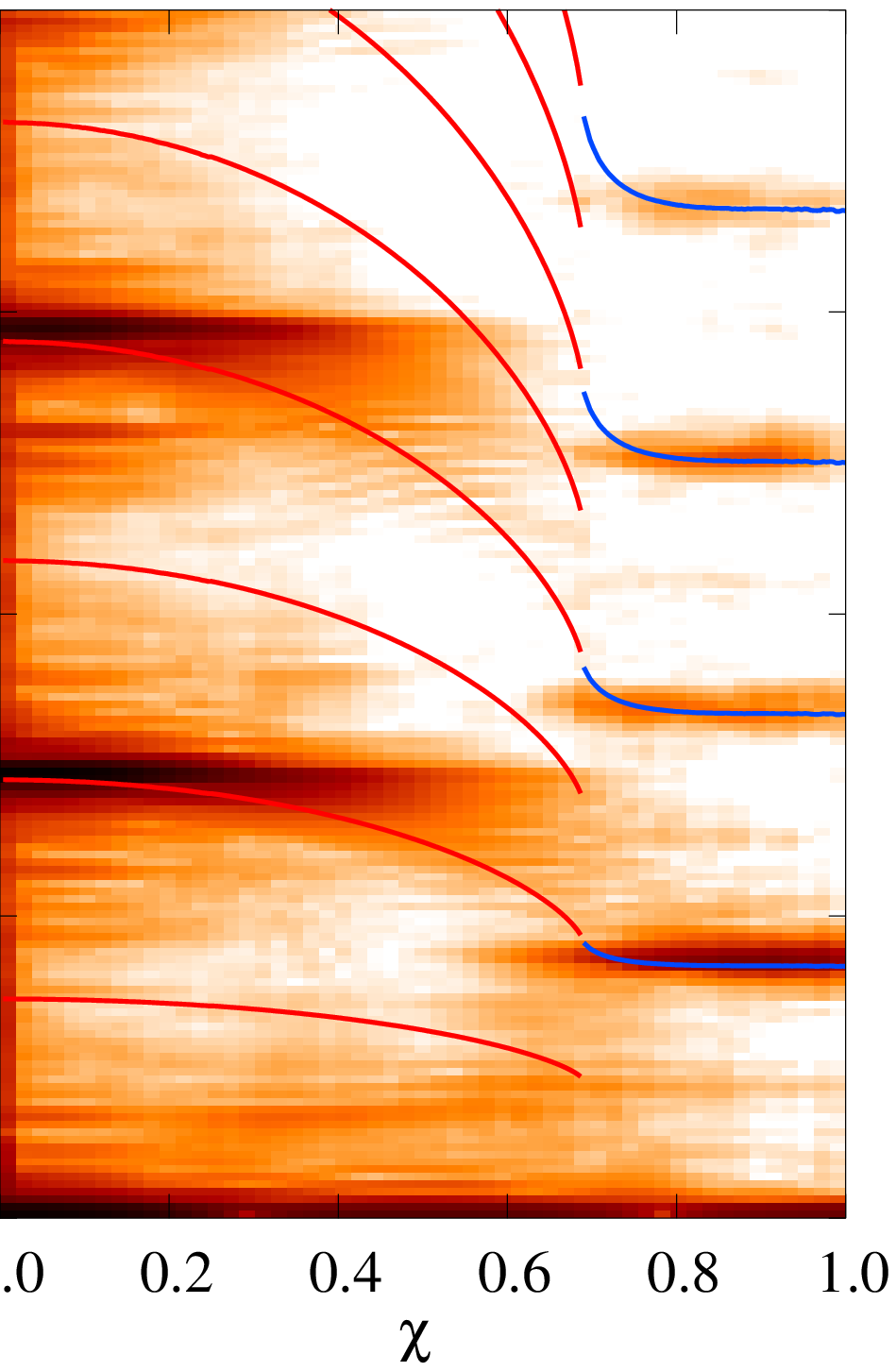}
  \caption{QPO continuum in the anelastic approximation. The logarithm of the effective amplitude,
  $<{\mathcal A} (\chi; f) >$, averaged along each individual magnetic field line, is shown 
in a frequency vs.  magnetic field line plot. The central and right panels correspond to 
antisymmetric and symmetric initial perturbations, respectively, while the left panel 
corresponds to a mixed initial perturbation with no symmetry. The (color) scale is the same 
in the three panels, ranging from a minimum value (white) to the maximum value (black). 
Solid lines show the continuum of standing-wave frequencies 
(repeated at different overtones), as obtained by our semi-analytic model,  
for open (red) and closed (blue) magnetic field lines inside the star.}
  \label{fig:mns2continuum}
\end{figure*}

We perform simulations using, initially, two different numerical resolutions, 
$n_r \times n_\theta=$ $80\times 40$ and $80\times 80$, where $n_r$ is the number 
of cells in the radial direction and
$n_\theta$ in the angular direction, for a full-grid setup ($\theta$ up to $180^o$).  We integrate in time for $800$~ms, which corresponds to about $40$ times the Alfv\'en crossing time, which is the dynamical timescale in our anelastic simulations.  As expected, the amplitude of oscillations is damped throughout
most regions in the star, due to the existence of a continuum of frequencies. At the same time, 
we observe persistent, long-lived QPOs near the pole and within the region of closed field
lines, as expected by our semi-analytic model and in agreement with ~\cite{Sotani2008}.

Since we do no expect global collective oscillations (normal modes) to be present, the analysis of the oscillations has to be performed locally. We denote by ${\mathcal A}(r,\theta; f)$ the 
amplitude of the Fourier transform of $\dot Y(r, \theta; t)$ at each point in the star, 
corresponding to an {\it effective oscillation amplitude} at a given frequency, for a given, 
finite, simulation time.  At the frequencies predicted by our semi-analytic model we find strong 
QPOs near the magnetic axis and within the region of closed magnetic field lines. For 
overtones, the number of nodes agrees also with the predictions of the semi-analytic model. 
Once we identify the QPOs, guided by the semi-analytic model, we refine the determination 
of the QPO frequencies by considering the maximum effective amplitude near the magnetic axis 
and within the region of closed magnetic field lines. 
The values of different extracted QPO frequencies (up to a certain overtone) are displayed
in Table~\ref{tab:freqs}, along with their relative difference when compared to the predictions
of the semi-analytic model. Although our anelastic approach and the semi-analytic model
employ very different approximations, the agreement in the QPO frequencies is remarkable. 
The frequency of the fundamental QPO agrees to within 4\% between the two approaches, 
while the agreement in all other frequencies is at the 1-2\% level.

Fig.~\ref{fig:mns2uqpo} displays the spatial pattern of the effective amplitude 
$\cal A$ for symmetric and antisymmetric ``upper'' QPOs, up to the third overtone.
Notice that  $\cal A$ has been rescaled by $r\sin\theta$ in order to correspond to a 
velocity component in a unit basis and not in the coordinate basis - in the latter
the amplitude is maximum exactly at the pole. For all overtones, the maximum
amplitude of the QPOs appears at the surface, at a location near the magnetic axis. 
As expected, the $U^{(-)}_n$ QPOs are antisymmetric w.r.t. the equatorial plane, 
while the $U^{(+)}_n$ QPOs are symmetric. The number of nodes along the 
magnetic axis and their precise location agrees with the prediction of the semi-analytic 
model (the nearly horizontal lines in this Figure are the predicted nodal lines).

Fig.~\ref{fig:mns2lqpo} displays the spatial pattern of the effective amplitude 
$\cal A$ (again rescaled by $r\sin\theta$) for the ``lower'' QPOs, up to the second overtone.
The amplitude of the ``lower'' QPOs is appreciable only within the region of closed
field lines (in the third panel of Fig.~\ref{fig:mns2lqpo}, a nonzero pattern near
the magnetic axis is a contamination of the finite FFT by other QPOs). As predicted
by our semi-analytic model, the fundamental QPO, $L_0$, has one nodal line crossing
the region of closed field lines, while each consecutive overtone has an additional
nodal line, each time added in a symmetric fashion. Since the initial perturbation used 
in this simulation had no particular symmetry w.r.t. the equatorial plane,
the precise location of the nodal lines inside the region of closed field lines 
cannot be easily be predicted by the semi-analytic model. The resulting patterns 
are nevertheless very close to those obtained for simulations that excite only QPOs 
of a particular symmetry, i.e. to $L^{(+)}_0$, $L^{(-)}_1$ and $L^{(+)}_2$.

Using the magnetic field structure of the unperturbed star we can map 
the time-derivative of the displacement,  $\dot Y (r, \theta; t)$, from polar 
coordinates to the magnetic-field-line-adapted coordinates, $\dot  Y (\xi, \chi; t)$. 
This allows us to define the absolute value of the maximum of $\dot Y$
along each magnetic field line, at any given time, i.e.  $| \dot  Y_{\rm max} (\chi; t)|$. 
The time-evolution of this quantity, for different magnetic field lines is shown in 
Fig.~\ref{fig:mns2evolution}, for the $80 \times 80$ reference model. Here, 
one can clearly observe how the initial perturbation 
evolves into long-lived QPOs near $\chi \sim 0$ and $\chi \sim 0.9$. In contrast, 
the oscillations at intermediate points are damped in time 
(see Sec.~\ref{sec:damping} for a discussion on the damping rates).

In a similar way, one can map the effective amplitude ${\mathcal A} (r, \theta; f)$ 
to  $ {\mathcal A} (\xi, \chi; f)$ and then compute its average
value along each magnetic field line, as
\begin{equation}
  <  {\mathcal A} (\chi; f) >=
  \int_0^1 d\xi \, {\mathcal A} (\xi, \chi; f).
\end{equation}
The left panel of Fig.~\ref{fig:mns2continuum} shows the distribution of  
$ < {\mathcal A} (\chi; f) >$ for different magnetic field lines and frequencies.
One can see that the effective amplitude displays local maxima at the predicted QPOs. 
In addition, a broader agreement with the continuum frequency parts for standing waves, 
predicted by the semi-analytic approach is also evident (red and blue lines overplotted).
Since no particular symmetry was present in the initial perturbation for this simulation, 
both symmetric and antisymmetric upper QPOs appear. In the middle and right panels  
we show similar plots, but for antisymmetric or symmetric initial data, respectively. This
time, only the corresponding  sub-families of upper QPOs appear. 
Moreover, the lower QPOs (near $\chi\sim 0.9$) appear at all integer multiples, since 
the frequencies of symmetric and anti-symmetric QPOs of same order coincide. Fig.~\ref{fig:mns2continuum} demonstrates that (at least for low amplitude perturbations) 
the QPOs excited by symmetric and antisymmetric perturbations completely  decouple. 
This may be relevant to the appearance  of only odd-integer multiples in the observed 
QPO frequencies in SGR 1806-20.

\subsection{Validity of the semi-analytic approach}
\label{validity}

The semi-analytic model of  Section~\ref{sec:alfven_oscillations} predicts that both
the upper QPOs and lower QPO overtones should appear at exact integer multiples 
of their corresponding fundamental frequency. The results presented in Table~\ref{tab:freqs} 
agree with this to within $2\%$, which is  comparable to the frequency resolution 
of the FFT, which was $1.25$~Hz. The only exception is the fundamental upper 
QPO frequency, which shows a larger difference w.r.t. the expected value, based on
the obtained values for the frequencies of the overtones, that cannot be explained 
by the finite frequency resolution of the FFT. We have checked that this is indeed 
not a numerical artifact, since the numerical frequency does not converge to the
semi-analytic result with increasing resolution.  As a result of this departure of
the fundamental upper QPO frequency from the prediction of exact integer multiples 
one can observe an interference of the first antisymmetric overtone, 
$U^{(-)}_1$, with  three times the frequency of the fundamental QPO, $U^{(-)}_0$, 
 which produces a beat frequency of
$f_{\rm beat} = (3\, U^{(-)}_0 - U^{(-)}_1) / 2$. 
Note that the frequency of the beat does not change appreciably with resolution. Measuring 
the beat frequency, the difference $(3\, U^{(-)}_0 - U^{(-)}_1)$ can be computed with 
higher accuracy than from the FFT itself. Using the last three "beats" in the simulation 
 we obtain a value of  $(3\, U^{(-)}_0 - U^{(-)}_1)=6.6$~Hz with a frequency resolution 
of $0.4$~Hz. The origin of this frequency difference with respect to the semi-analytic model
is the short-wavelength assumption in the latter.  However, for the fundamental upper QPO 
the wavelength of the corresponding standing-wave is twice the size of the star itself, and 
therefore our approximation becomes inaccurate. For all other overtones, however, the
short-wavelength approximation is valid with good accuracy.

\begin{figure}
  \centering
  \resizebox{0.45\textwidth}{!}{\includegraphics*{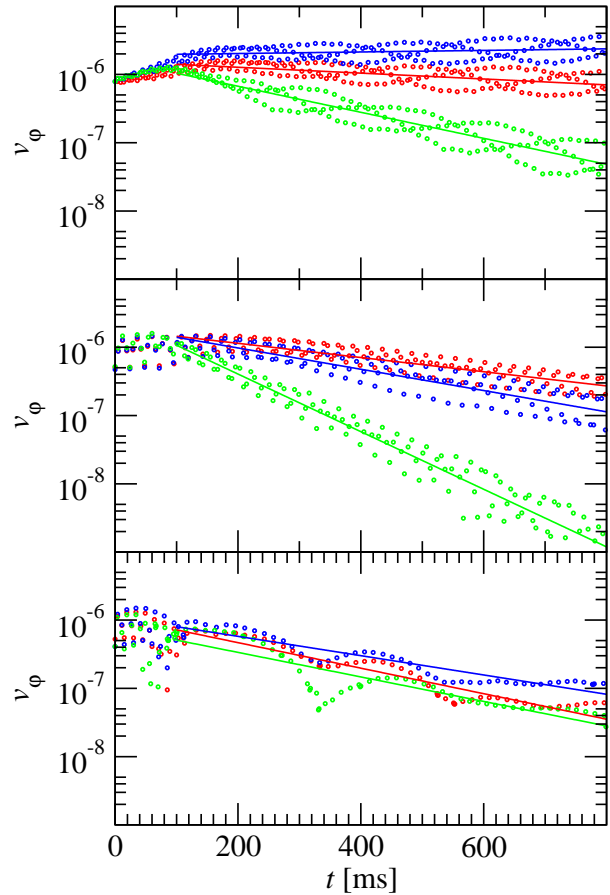}}
  \caption{
Time evolution of the maximum absolute value of $v_\varphi$ during
oscillations, at three representative points inside the star: at the
points where the upper QPO (upper panel ) and lower QPO (middle panel) 
show their respective maximum effective amplitudes  and at 
$r=0.65 r_e$ and $\theta=\pi/4$ (lower panel), a point on an open field line, 
far from both regions where the upper and lower QPOs are prominent. 
Three resolutions were used:, $80\times 20$ (green dots), $80\times 40$ (blue dots)
 and $160 \times 40$ (red dots). Straight lines are fits to  an exponential law 
for $t>100$~ms.}
  \label{fig:mns2spec}
\end{figure}

\subsection{Dependence on the symmetry of the initial perturbations}

In order to check the dependence of our numerical simulations on the symmetry of the initial perturbation we have performed two more simulations of model MNS2 with a symmetric ($a_{2}=0$ and $a_3=10^{-5}$) and an antisymmetric perturbation ($a_{2}=10^{-5}$ and $a_3=0$).  Since the background model is symmetric with respect to the equatorial plane and the perturbations are either symmetric or antisymmetric, we can perform the simulations using only one quadrant ($90^o$), and imposing the appropriate symmetries in the equatorial plane depending on the initial perturbation. Therefore, the boundary conditions for the perturbed quantities, $v^\varphi$ and $B^\varphi$, in the equatorial plane, depend on the symmetry of the perturbation with respect to the equator: for symmetric perturbations, $\partial_\theta v^\varphi=0$ and $B^\varphi=0$, and for antisymmetric perturbations, $v^\varphi=0$ and $\partial_\theta B^\varphi=0$.  We have performed simulations using $80\times 40$ and $80\times20$ grid cells in $(r\times \theta)$, which correspond to the same resolution as the full $180^o$ simulations  of $80\times 80$ and $80\times 40$ cells respectively. We have also performed low resolution simulations of  $80\times 40$ cells in a $180^o$ domain, checking that there is complete agreement with the $90^o$ simulations with the same effective resolution of $80\times 20$ cells.

When using either symmetric or antisymmetric initial perturbations, the spatial pattern at each QPO frequency is similar to the corresponding cases in Fig.~\ref{fig:mns2uqpo}, and the number of nodes and their location match as well (see also central and right panels of Fig.~\ref{fig:mns2continuum}). The location of the nodes of the lower QPOs can be predicted because of the particular symmetry of the perturbations. Similarly, for the lower QPOs, the spatial pattern for each QPO frequency agrees in the number of nodes and their location with the corresponding semi-analytic model, for either symmetric or antisymmetric initial perturbations. A $90^o$ spatial phase difference is found between symmetric and antisymmetric perturbations, which is expected for the type of initial data we use.

We note that the time evolution of the perturbations in all the different background models we use is qualitatively similar to the results reported using the reference model.

\subsection{Damping rates and viscosity}
\label{sec:damping}

According to our analysis of section~\ref{sec:alfven_oscillations}, long-lived QPOs (upper and lower) are expected near the magnetic axis and inside the region of closed field lines, corresponding to extrema in continuum parts of the spectrum. In other regions, initial oscillations should be damped quickly, due to phase mixing. Apart from this, our simulations are affected by intrinsic numerical viscosity (due to finite-differencing) which produces an additional damping of the oscillations everywhere in the star. We use our reference model MNS2, with an antisymmetric initial perturbation, in order to examine the effect of numerical resolution on the damping rate of oscillations at different regions inside the star, i.e.~in order to examine the effect of numerical viscosity. We compute the evolution of this model with low resolution ($80\times 20$), standard resolution ($80\times 40$) and high resolution ($160\times 40$), for a half-grid setup. Fig.~\ref{fig:mns2spec} shows the time evolution of the maximum absolute value of $v_\varphi$ during the oscillations, at three representative points inside the star: at the points where the upper QPO (upper panel ) and lower QPO (middle panel) show their respective maximum effective amplitudes,  and at a point on an open field line, but far from the regions where both the upper and lower QPOs are prominent (lower panel), specifically, at $(r,\theta)= (0.65 r_e,\pi/4)$.

For the two QPO families (upper and middle panels of  Fig.~\ref{fig:mns2spec}) we observe that the damping is significantly reduced at high resolution. Especially for the upper QPOs (upper panel), already $40$ angular cells seem to be sufficient to have practically negligible numerical damping. In contrast, the lower QPOs are always damped in time, even with the highest resolution we use. This could be due, in part,  to the fact that closed magnetic field lines are much shorter than open field lines, and therefore much higher resolution is needed to  get a convergent result. From these results it is difficult to conclude what the actual damping of the lower QPOs would be, as even higher  resolution simulations would be necessary to further diminish the numerical viscosity. 
Finally, we observe a very strong damping rate for oscillations at the point which is far from the two regions where the upper or lower QPOs are prominent (lower panel of Fig.~\ref{fig:mns2spec}), and this rate does not decrease appreciably with increasing resolution. This indicates that the damping at such points must be due to phase mixing.

In some magnetohydrodynamic systems, quasi-modes can be
found, which are normal modes with a very large imaginary
part in their frequency (see \cite{Levin2007} and  references therein). 
We have investigated the relation of the QPOs found
here to possible quasi-modes in the limit of high viscosity, by adding artificial
viscosity to our simulations. In particular, we performed a simulation of 
the reference model MNS2, with a resolution of $80\times 20$ cells (in a half-grid setup) 
and an initial antisymmetric perturbation, adding artificial viscosity by means 
of a Kreiss-Oliger 4th order term in the $\varphi$ momentum equation \citep{KO1973}. 
We used a large pre-factor in this viscosity term equal to $0.1$. In this case we 
indeed find results that are consistent with global collective oscillations with a
large imaginary part (quasi-modes). The computed real parts of quasi-mode 
frequencies and their phase show a minimal variation for different 
magnetic field lines ($\chi$) while the real part of the frequency of quasi-modes 
is equal to the real part of the frequency of the corresponding QPOs 
in the non-viscous case. 

\begin{table*}
  \begin{minipage}{0.63\textwidth}
  \caption{Frequencies (in Hz) of the lower and upper QPOs for all models. Values
    computed with the semi-analytic standing-wave approximation are labeled "sa",
    numerical values of the fundamental frequency obtained in the anelastic 
    approximation are labeled "num" and fitted values using the harmonics are labeled 
    ``fit'' (see text for details). In parentheses we show the relative difference with
    respect to the corresponding semi-analytic value.}
  \label{tab:freqsall}
  \begin{tabular}{llllllllllll}
    \hline
    Model &  $U_0^{(-)\rm sa}$ & $U_0^{(-)\rm num}$ & $U_n^{(-)\rm fit}/(2n+1)$ &  $L_0^{\rm sa}$ & $L_0^{\rm num}$ & $L_n^{\rm fit}/(n+1)$ \\[0.3 em]  
    \hline
    MNS1  &  3.63             & 3.50  (3.6\%)   & 3.62  (0.3\%)       & 4.17           & 4.12  (1.2\%) & 4.23  (1.4\%) \\[0.3 em] 
    MNS2  &  36.3             & 35.1  (3.3\%)   & 36.8  (1.4\%)       & 41.7           & 41.3  (1.0\%) & 42.16 (1.1\%) \\[0.3 em] 
    MNS3  &  91.2             & 87.4  (4.2\%)   & 89.2  (2.2\%)       & 104.7          & 103.1 (1.5\%) & 106.2 (1.4\%) \\[0.3 em]  
    HMNS2 &  35.4             & 33.7  (5.0\%)   & 35.9  (1.5\%)       & 43.3           & 43.73 (1.0\%) & 43.4  (0.2\%) \\[0.3 em] 
    LMNS2 &  35.2             & 33.7  (4.2\%)   & 35.0  (0.6\%)       & 39.2           & 38.71 (1.2\%) & 39.51 (0.8\%) \\[0.3 em] 
    S1    &  21.7             & 20.0  (7.8\%)   & 21.96 (1.2\%)       & 25.4           & 25.95 (2.2\%) & 25.9  (2.0\%) \\[0.3 em] 
  \hline
  \end{tabular}
  \end{minipage}
\end{table*} 

\subsection{Empirical relations}
\label{sec:III}

 In other to check the dependence of the QPO frequencies on the magnetic field strength and on the mass of the neutron star we also compute the evolution of models for which we have varied either the central current density (models MNS1 and MNS3) or the central density (models HMNS2 and LMNS2), with respect to the reference model MNS2. The equilibrium properties of these models are shown in Table~\ref{tab:ini_models}. The simulations are performed in a half-grid setup with a resolution of $80\times 20$ cells. Our results are summarized in Table \ref{tab:freqsall}. Frequencies computed with the semi-analytic, short-wavelength
approximation are labeled "sa", while those obtained in the anelastic approximation are labeled "num".
In parentheses we show the relative difference with respect to the corresponding semi-analytic value. 
As for the reference model MNS2, we observe a relative difference of several percent for the 
frequency of the fundamental upper QPO. As we explained in Sec. \ref{validity}, this is because the 
wavelength of the fundamental QPO is too large for the short-wavelength approximation to be
more accurate. However, since frequencies of all overtones show a relative difference of order 1-2\% 
only, with respect to the corresponding frequencies of the semi-analytic model, if we assume that
these are exact integer multiples of a fundamental frequency, then we can construct a fit that yields
a value for the fundamental frequency compatible with this assumption. These values are labeled
as ``fit'' in Table \ref{tab:freqsall}. 

When varying the strength of the magnetic field we find that, as expected from the results of~\cite{Sotani2008}, the QPO frequencies change linearly with $B$. Also, the effect of changing the stellar mass is consistent (within a few percent) with the expansion in terms of the compactness, ($M/R$), used in the empirical relations presented by~\cite{Sotani2008}. This agreement allows us to construct empirical relations for all QPO frequencies, taking advantage of the ($M/R$) expansion in~\cite{Sotani2008}, which was constructed for a set of different EOS (while here we only consider a polytropic EOS). We note that in practice, the larger number of data points used in~\cite{Sotani2008} results in a more accurate empirical relation than using the few models presented here. In order to construct new empirical relations, we use the frequencies labeled ``fit'' in Table~\ref{tab:freqsall}. 

Our empirical relations (constructed as described above) are 
\begin{eqnarray}
 f_L &=& 56.8 (n+1) \left[1-4.55\left(\frac{M}{R}\right)+6.12 \left(\frac{M}{R}\right)^2 \right ]
             \nonumber \\ 
   && \times \left(\frac{B}{4\times 10^{15}{\rm G}} \right) \quad ({\rm Hz}),
\end{eqnarray}
for the family of lower QPOs and 
\begin{eqnarray}
 f_U &=& 48.9 (n+1) \left[1-4.55\left(\frac{M}{R}\right)+6.12 \left(\frac{M}{R}\right)^2 \right ]
             \nonumber \\ 
   && \times \left(\frac{B}{4\times 10^{15}{\rm G}} \right) \quad ({\rm Hz}),
\label{empU}
\end{eqnarray}
for the family of upper QPOs. Here, $n$ indicates the order of the QPO, with $n=0$ corresponding
to the fundamental QPO. Notice that the above two relations include all obtained QPOs, irrespective 
of their symmetry with respect to the equatorial plane. For the lower QPOs, the symmetric and 
antisymmetric cases have coincident frequencies and appear at all integer multiples. For the 
upper QPOs, the antisymmetric cases correspond to $n=0,2,4,..$ (odd integer multiples) while 
the symmetric cases correspond to $n=1,3,5,..$ (even integer multiples). Such empirical relations 
are useful for trying to interpret observed QPO frequencies and for extracting useful information 
about the  characteristics of the source (see the discussion in Sec. \ref{sec:discussion}). 

We can now apply the above empirical relations to the two known SGR sources that exhibit the
QPO behavior in their X-ray tail. For this, we use the set of equilibrium models presented in ~\cite{Sotani2007a}, which were constructed for different tabulated EOS and different masses. 
Within this representative set of models, the compactness ratio $M/R$ ranges from 0.14 to 0.28.
If we assume that the observed frequencies of 30, 92 and 150Hz in SGR 1806-20 are produced 
as QPOs near the magnetic poles (upper QPOs) and are odd-integer multiples of the fundamental 
frequency of 30Hz, then the empirical relation (\ref{empU}) restricts the dipolar component
of the magnetic field to be in the range of $5\times10^{15}$G to $1.2\times10^{16}$G. 
Similarly, if the observed frequencies of 28, 53, 84 and 155Hz in 
SGR 1900+14 are produced as QPOs near the magnetic poles (upper QPOs) and 
are near-integer multiples of the fundamental frequency of 28Hz, then the dipolar 
component of the magnetic field is restricted to be in the range of $4.7\times10^{15}$G to 
$1.12\times10^{16}$G. 

Notice that the above values refer to the strength of the magnetic field
at the poles. For the particular form of the dipolar magnetic field, the mean value at the surface
is a factor of roughly 2/3 smaller than the strength at the pole. Therefore, for the chosen sample
of equilibrium models and taking both SGR sources into account, the {\it mean surface magnetic
field strength} of the dipolar component is restricted to be in the range of $3\times10^{15}$G to 
$8\times10^{15}$G, if the fundamental QPO frequencies are identified as above. 
Still, the fundamental QPO frequencies could be any integer fraction of 30Hz in 
SGR 1806-20 or of 28Hz in SGR 1900+14. For this reason, the range determined above  is only
an {\it upper limit} to the possible strength of the magnetic field. Notice that using the observed 
period and spin-down rate of the magnetar in the two SGR sources, one can estimate the mean 
surface magnetic field to be about $6\times10^{14}$G for SGR 1900+14 and $8\times10^{14}$G 
for SGR 1806-20, which is within the upper limit established above.

\section{Discussion}
\label{sec:discussion}

We have presented the first two-dimensional numerical simulations of axisymmetric, 
torsional Alfv\'{e}n oscillations in magnetars in the anelastic approximation, in which fluid modes 
are suppressed.  In addition to the numerical simulations, we have also computed Alfv\'{e}n oscillation 
frequencies along individual magnetic field lines with a semi-analytic approach, employing a 
short-wavelength approximation. The continuum of standing-wave oscillations obtained using 
the semi-analytic approach agrees remarkably well with QPOs obtained via our 
two-dimensional simulations This agreement will allow for a comprehensive study 
of Alfv\'{e}n QPOs for a large number of different models, without the need of time-consuming 
simulations. 

Our semi-analytic approach is not restricted to a dipolar field, but can be extended to 
other magnetic field configurations, while our  anelastic approximation does not suffer from 
the numerical instabilities encountered in the linear study of \cite{Sotani2008}. Furthermore,  our anelastic approach could be extended to include shear waves in the crust, which may prove to be 
important for the interpretation of observed QPOs. 

Our results agree qualitatively with those obtained 
by~\cite{Sotani2008}, who worked in the linear regime, but otherwise used a similar computational 
setup as the one employed here. One major difference is our finding that apart from the anti-symmetric upper QPOs, symmetric upper QPOs also exist, which completes the description of the possible QPO families. In addition, our implementation of the boundary condition at the surface, motivated by the presence of a solid crust in real magnetars, results in a somewhat different spectrum than in~\cite{Sotani2008}. Specifically, QPOs generated 
near the axis have a nonzero amplitude at the surface, which would allow these oscillations to reach 
the magnetosphere, while in ~\cite{Sotani2008} the oscillations nearly vanished at the surface.
This difference in the applied boundary conditions at the surface results in a different structure of 
standing-wave QPOs near the pole, hence the quantitative difference between our empirical relations 
and those given in~\cite{Sotani2008}. In contrast, the QPOs produced in the region of closed magnetic
field lines are not affected by the boundary condition at the surface. 

Improvements of the Alfv\'en QPO model can be achieved by the inclusion of a realistic crust in the 
nonlinear simulations and of correspondingly more realistic boundary conditions at the crust-core 
interface and at the stellar surface. The crust is expected to oscillate predominantly at the frequencies
of the turning-point QPOs, since the magnetized core will pull at the crust coherently at these frequencies, 
with an amplitude that is only slowly reduced as $1/t^{1/2}$ \cite{Levin2007}. In addition, the 
crustal torsional normal mode oscillations may also be triggered. Regarding the edge-QPO that is 
present in our semi-analytic model for antisymmetric standing waves, at $\chi\sim 0.66$ (at the
last open magnetic field line), even though we could not detect the presence of a corresponding
QPO in our two-dimensional simulations in the anelastic approximation, the situation may change
in the presence of a crust and with a corresponding realistic boundary condition that will couple 
magnetic field lines in this area. It is thus possible that the 18 and 26Hz QPOs in SGR 1806-20
are due to crust oscillations and/or edge QPOs, but additional turning-point QPOs, due to a more
complicated magnetic field structure than considered here, could also be expected. The
polar-parity Alfv\'en oscillations are also of particular interest, as these could
be excited after an SGR event, in addition to the torsional modes studied here (see \cite{Sotani2009} for a recent study of polar Alfv\'en
 oscillations in relativistic magnetar models). 

Other effect that has been neglected so far in previous works, as well
as in our simulations is the presence of a superfluid phase in the
magnetar interior. Recently \cite{Andersson08} have suggested that the
effects of including this more realistic kind of matter should
influence the frequencies by a factor of a few and could potentially have an
impact on the computation of the QPO frequencies. If this were the case,
the frequencies computed in this paper and the upper limits for the
magnetic fields proposed could change by a factor of a few.

Finally, the coupling to an exterior magnetosphere has to be taken into account, as well as the 
precise mechanism by which the X-ray flux in the afterglow of SGR flares 
is modulated by Alfv\'en oscillations. We are planning to address these issues in future extension of the
work presented here.

\section*{Acknowledgments}

It is a pleasure to thank Kostas Kokkotas for a continuous exchange of 
results and ideas, while this work
was completed in parallel to \cite{Colauida2009}. We also thank Luciano 
Rezzolla for helpful discussions. 
We are particularly grateful to Hajime Sotani for performing simulations 
with the aim of a direct comparison  
between the linear and nonlinear codes. This work was supported  by the 
GSRT Greece-Spain bilateral grant, the DAAD Germany-Greece bilateral grant, 
the EU network ILIAS,  the Spanish Ministry of Education and Science 
(AYA 2007-67626-C03-01), and the Collaborative Research Center on 
{\it Gravitational Wave Astronomy} of the Deutsche Forschungsgesellschaft 
(DFG SFB/Transregio 7).


\end{document}